\documentclass[11pt]{article}
\usepackage{lineno}
\usepackage{multirow}
\usepackage{amsmath}
\usepackage{float}
\usepackage{graphicx} 
\usepackage{amssymb,amscd,amsmath,amsthm,color}
\usepackage[T1]{fontenc} 
\usepackage{subcaption}
\usepackage{enumitem}
\usepackage{mathtools}
\usepackage{tcolorbox}
\usepackage{verbatim}
\usepackage{mathrsfs}
\usepackage{physics}
\usepackage{mathtools}
\usepackage{slashed}
\usepackage{longtable}
\usepackage{jheppub}
\usepackage{enumerate}
\usepackage{hyperref}
\usepackage{ulem}
\newcommand{\Slash}[1]{\not\!#1}

\numberwithin{equation}{section}

\def\cH{\mathcal{H}}

\def\cD{\mathcal{D}}

\def\<{\langle}
\def\>{\rangle}

\def\be{\begin{equation}}
\def\ee{\end{equation}}

\def\cO{\mathcal{O}}
\def\cN{\mathcal{N}}
\def\cL{\mathcal{L}}

\makeatletter
\g@addto@macro\bfseries{\boldmath}\g@addto@macro\bfseries{\boldmath}
\makeatother
\title{\boldmath Only Flat Spacetime is Full BPS in Four Dimensional $\mathcal{N}=3$ and $\mathcal{N}=4$ Supergravity}

\author[a]{Abhinava Bhattacharjee,}
\emailAdd{abhinava19@iisertvm.ac.in}
\affiliation[a]{School of Physics, Indian Institute of Science Education and Research Thiruvananthapuram,\\
Thiruvananthapuram 695551, India}

\author[b]{Subramanya Hegde,}
\emailAdd{subbu@mpp.mpg.de}
\affiliation[b]{Max-Planck-Institut f\"ur Physik, Werner-Heisenberg-Institut, Boltzmannstr. 8, 85748 Garching bei M\"unchen, Germany}
	
\author[a]{Bindusar Sahoo}
\emailAdd{bsahoo@iisertvm.ac.in}

\abstract{We investigate the fully supersymmetric solutions in a class of $\mathcal{N}=3$ and $\mathcal{N}=4$ higher derivative Poincar\'e supergravity theories. These class of theories are obtained within the framework of conformal supergravity using the standard Weyl multiplet and contains a set of terms related to the Weyl square term by supersymmetry. We work in the superconformal formalism and show that flat spacetime is the only fully supersymmetric solution in these theories. However, in  $\mathcal{N}=2$ Poincar\'e supergravity theories that falls into the same class, two fully supersymmetric stationary solutions exist: Bertotti-Robinson geometry ($AdS_2\times S^2$) and flat spacetime, which are known in the literature. We discuss the reason behind the richer vacua in $\cN=2$ supergravity compared to its $\cN=3$ and $\cN=4$ cousins.} 

\makeatletter
\gdef\@fpheader{}
\makeatother
\begin{document}
\allowdisplaybreaks
\maketitle
\flushbottom
\section{Introduction}
\label{sec:intro} Over the last four decades, supersymmetric solutions of supergravity theories have proven crucial for understanding various aspects of string theory and its implications on black hole physics. Supersymmetric soliton solutions played a central role in unraveling the underlying hidden dualities between the different string theories \cite{Hull:1994ys} and understanding the microscopic origin of Bekenstein-Hawking entropy in string theory \cite{Strominger:1996sh}. Notably, in $\mathcal{N}=2$ supergravity, horizons of the extremal black holes are fully supersymmetric, and using this fact, a macroscopic entropy formula was derived, which facilitated a successful comparison with the microscopic state counting in various $\mathcal{N}=2$ string compactifications \cite{Mohaupt:2000mj}. 

The $\mathcal{N}$-extended supergravity\footnote{When we say $\mathcal{N}$-extended supergravity, we mean ungauged $\mathcal{N}$-extended Poincar\'e supergravity throughout the paper.}
theories are invariant under diffeomorphism, local Lorentz transformation as well as local supersymmetry with $\mathcal{N}$ arbitrary spinor parameters $\epsilon^i(x)$, where $i$ runs from $1$ to $\cN$. Let us collectively denote the bosons and fermions in the theory as $\{B(x)\}$ and $\{F(x)\}$ respectively. A solution is said to be supersymmetric, if there exists a subset $\{\epsilon^j(x)\}$ of local supersymmetries such that 
\begin{equation}\label{KSE}
\delta_{\epsilon^j(x)}\{B(x)\}=0,~~\delta_{\epsilon^j(x)}\{F(x)\}=0, ~~j=1,\dots,M\leq \mathcal{N}
\end{equation}
The spinors in the subset are known as Killing spinors and the schematic equations in \eqref{KSE} are the Killing Spinor equations. For classical field configurations, the fermions vanish, and the first condition of $\eqref{KSE}$ is trivially satisfied. A fully supersymmetric solution supports the maximal number of Killing spinors, i.e., $M=\mathcal{N}$ ( $4 
 \mathcal{N}$ conserved supercharges).\footnote{In four spacetime dimensions, the basic spinors are Majorana spinors with four real components. For spinors in other dimensions, look at \cite{Freedman:2012zz}. } For $M< \mathcal{N}$, the solution is said to preserve $M/\mathcal{N}$ of total supersymmetries. The Killing spinor equations serve a dual purpose. Firstly, for a given solution, one can find the Killing spinors by solving the vanishing gravitini field variations
 \begin{equation}
     \delta \psi^{i}_{\mu}= \mathscr{D}_{\mu}\epsilon^i(x)+\dots=0,
 \end{equation}which are first order differential equations in the spinor parameters and the variation of the other fermions are linear in the spinor parameters. The form of the supercovariant derivative $\mathscr{D}_{\mu}$ depends on the details of the supergravity theory. The integrability conditions of these equations are ensured from the vanishing variation of the supercovariant gravitino field strength
 \begin{equation}
     \delta R_{\mu\nu}{}^{i}= \mathscr{D}_{[\mu}\mathscr{D}_{\nu]}\epsilon^{i}(x)+\dots =0
\end{equation}
Secondly, assuming the existence of the Killing spinors, one can consider \eqref{KSE}  as conditions imposed on the fields and find the field configurations that satisfy the conditions. To obtain $M/\mathcal{N}$-supersymmetric solutions, one must impose appropriate embedding conditions to make $\mathcal{N}-M$ Killing spinors dependent. Minkowski space is always a maximally supersymmetric solution to any $\mathcal{N}$-extended supergravity in any spacetime dimensions possessing $\mathcal{N}$ constant Killing spinors, but various other maximally supersymmetric solutions exist in different supergravity theories in diverse dimensions.

The first classification of supersymmetric solutions using Killing spinor equations was performed by Tod in \cite{Tod:1983pm} in the context of pure $\mathcal{N}=2$ two-derivative supergravity. The fully supersymmetric solutions are Bertotti-Robinson geometry $(AdS_2\times S^2)$, Minkowski space and pp-wave spacetimes \cite{Kowalski-Glikman:1985fjy, LopesCardoso:1998tkj}. In \cite{Mohaupt:2000mj, LopesCardoso:2000qm}, the authors considered a class of higher derivative $\mathcal{N}=2$ theories where the higher derivative corrections are encoded in a holomorphic prepotential function and showed that the only $\mathcal
{N}=2$ fully supersymmetric stationary solutions are $AdS_2\times S^2$ and flat spacetime. In this theory, the extremal black holes, which preserves $1/2$ of the supersymmetries, are of the form of Israel-Wilson-Perj\'es (IWP) metric and it approaches $AdS_2\times S^2$ geometry at the horizon and Minkowski space at asymptotic infinity \cite{LopesCardoso:2000qm}. The enhancement of supersymmetry at the horizon led to the discovery of attractor mechanism in the context of $\mathcal{N}=2$ black holes \cite{Ferrara:1996dd}. For dimension $d> 4$, a plethora of solutions appear in supergravity theories with $8$ supercharges, that include black holes, black rings, black branes, Godel type metric, etc. \cite{Gauntlett:2002nw, Gutowski:2003rg, Chamseddine:2003yy, Figueroa-OFarrill:2002ecq}. 

For $\mathcal{N}=4$, $d=4$ supergravity, supersymmetric solutions
 include the Super-Israel-Wilson-Perj\'es (SIWP) solutions and pp waves spacetime, but none of these solutions are fully supersymmetric \cite{Tod:1995jf, Bergshoeff:1996gg, Alonso-Alberca:2002efr, Bellorin:2005zc}. From a string theory point of view, $\mathcal{N}=4$ supergravity can be obtained from compactification of $\mathcal{N}=1, d=10$ heterotic string theory on  $T^6$ or compactification of $\mathcal{N}=2, d=10$ type-IIA string theory on $K3\times T^2$ \cite{Chamseddine:1980cp}. These parent theories have a flat Minkowski space as the only maximally supersymmetric solution \cite{Figueroa-OFarrill:2002ecq}. Furthermore, it has recently been argued in \cite{Chen:2024gmc} that $AdS_2\times S^2$, which typically appears as a near horizon geometry of extremal supersymmetric black holes, may not possess four Killing spinors (16 supersymmetries). In four dimensions, no solution except flat spacetime has been found to preserve all the 16 supersymmetries. A simple criterion to distinguish which two-derivative ungauged supergravity admit fully supersymmetric non-flat solutions is given in \cite{Louis:2016tnz}. This criterion may explain why flat spacetime is the only fully supersymmetric solution in two derivative ungauged $\mathcal{N}=4$ supergravity in four dimensions. In this paper, however we prove that flat spacetime remains the only fully supersymmetric solution in four dimensional $\mathcal{N}=4$ supergravity in the presence of a class of higher derivative corrections that is obtained within the framework of conformal supergravity using the ``standard Weyl multiplet'' (see our discussion on the superconformal formalism later in this section for a definition of ``standard Weyl multiplet'') and contains all terms related to the Weyl squared terms by supersymmetry. Note that these terms are of particular interest in the context of macroscopic matching of $1/2-$BPS Dabholkar-Harvery states \cite{Dabholkar:1989jt} with the entropy of a $1/2-$BPS blackhole in higher derivative $\mathcal{N}=4$ supergravity \cite{Hubeny:2004ji,Dabholkar:2004dq, Sen:2004dp}. The existence of such a resolution is dependent on the existence of $AdS_2\times S^2$ solutions that preserve 16 supercharges in $\mathcal{N}=4$ higher derivative supergravity with the kind of higher derivative terms considered in this paper\footnote{Recently, in \cite{Chen:2024gmc}, an analysis of possible superconformal groups in the near horizon region was used to argue that such a decoupled near horizon $AdS_2\times S^2$ is not possible with 16 conserved supercharges.}. We also show that flat spacetime remains the only maximally supersymmetric solution even for $\cN=3$ supergravity in the presence of the same class of higher derivative corrections. We also discuss the reason behind the completely different structure of supersymmetric vacua for $\cN=4$ and $\cN=3$ supergravity in contrast with $\cN=2$ supergravity which admits richer vacua.
We work in the framework of conformal supergravity. Conformal supergravity \cite{Fradkin:1985am} is a supersymmetric extension of conformal gravity or, equivalently, the conformal extension of Poincar\'e supergravity. These theories possess a large number of local symmetries, namely diffeomorphism, local Lorentz transformations ($M$), dilatations ($D$), special conformal transformations ($K$), R-symmetries, ordinary supersymmetry (which we will often refer to as $Q$-supersymmetry), and special supersymmetry (often referred to as $S$-supersymmetry). A brief outline of the superconformal formalism is as follows \cite{Freedman:2012zz}:
\begin{itemize}
    \item One starts with a gauge theory of the corresponding superconformal group and applies appropriate curvature constraints to make it a theory of gravity.
    \item Consequently, some of the gauge fields become dependent. 
    \item  To close the algebra off-shell, one needs to introduce extra covariant fields, which together with the independent gauge fields form the so-called Weyl multiplet. The superconformal algebra is modified to a soft superconformal algebra where the structure constants become structure functions which depend on the extra covariant matter fields. These extra covariant fields are sometimes referred to as ``auxiliary fields''. \footnote{\label{footnote}There are different ways of introducing the auxiliary fields which leads to different variants of Weyl multiplets: Standard and Dilaton. In this paper we confine ourselves to the standard Weyl multiplet.}
    \item 
    Apart from the Weyl multiplet, one needs to introduce extra matter multiplets such as vector multiplet, hypermultiplet etc, which form representations of the soft superconformal algebra introduced above. The superconformal algebra will close off-shell on some of these multiplets and they are referred to as off-shell multiplets and on some other multiplets the superconformal algebra will close upon using their field equations. Such multiplets are referred to as on-shell multiplets.
    \item 
     One constructs a superconformally invariant action for these multiplets (both Weyl and matter) using various known methods in the literature. See for example, chiral density formula \cite{deRoo:1980mm, Mohaupt:2000mj} or vector-tensor density formula \cite{deWit:1982na, Claus:1997fk, deWit:2006gn} for construction of actions in $\cN=2$ conformal supergravity. For $\cN=4$ conformal supergravity a more general action principle based on covariant superforms was introduced in \cite{Butter:2016mtk, Butter:2019edc} which was also used in $\cN=3$ conformal supergravity \cite{Hegde:2022wnb} and $\cN=2$ conformal supergravity \cite{Hegde:2019ioy}.
     \item 
     A subset of the matter multiplets are used as compensating multiplets to transition from conformal to Poincar{\'e} supergravity \cite{Kaku:1978ea, deWit:1979dzm, Kaku:1978nz,deWit:1982na, Freedman:2012zz}. These subset of matter multiplets have the right number of degrees of freedom to compensate for the additional symmetries necessary to transition from conformal to Poincar{\'e} supergravity.
     \item
     Additionally, some of the auxiliary fields from the Weyl multiplet can be removed by using their field equations if one wants the theory purely in terms of the physical fields.
     \item 
     If one wants to have a higher derivative matter coupled Poincar{\'e} supergravity, one needs to take into account the action for the Weyl multiplet which will modify the field equations of the auxiliary fields and removing the auxiliary fields would generate a derivative expansion of the action in terms of the physical fields \cite{Butter:2016mtk, Butter:2019edc, Hegde:2022wnb}.
\end{itemize} 
This formalism has several technical advantages. Including higher derivative corrections to Poincar\'e supergravity is technically a complex problem. The off-shell formalism of conformal supergravity provides a systematic way to incorporate higher derivative corrections into Poincar\'e supergravity. The supersymmetry transformations are much simpler in the superconformal formalism and easier to deal with. Hence, we analyze the Killing spinors equations in the superconformal formalism, and at the very end, we impose the gauge fixing conditions to obtain the field configurations in the Poincar\'e supergravity theory.

This paper is organized as follows. In section \ref{N4sugra}, we briefly review $\mathcal{N}=4$ supergravity using the framework of conformal supergravity. In section \ref{N4susy}, we analyze the Killing spinor equations in the superconformal formalism and obtain the fully supersymmetric $\mathcal{N}=4$ field solution. In section \ref{N3sugra}, we review $\cN=3$ supergravity using the framework of conformal supergravity. In Section \ref{N3susy}, we obtain the fully supersymmetric $\cN=3$ solution by solving the Killing spinor equations in the superconformal formalism. In section \ref{N32}, we discuss the reason behind the existence of multiple vacua in $\cN=2$ supergravity in contrast to $\cN=4$ or $\cN=3$ supergravity. We conclude the paper with some future directions.

\section{\texorpdfstring{$\cN=4$ Higher Derivative Supergravity}{N=4 Higher Derivative Supergravity}}
\label{N4sugra}
In this section, we review four dimensional matter-coupled higher derivative $\mathcal{N}=4$ supergravity using the framework of conformal supergravity. For a detailed construction of $\cN=4$ conformal supergravity as well as the construction of matter coupled $\cN=4$ supergravity using the superconformal framework, see \cite{Bergshoeff:1980is,deRoo:1984zyh,Ciceri:2015qpa, Butter:2016mtk,Butter:2019edc}. The relevant multiplets are the $\cN=4$ Weyl multiplet and the $\cN=4$ vector multiplet. Among the vector multiplets, there are a fixed number of ``compensating multiplets'' that provide the necessary degrees of freedom to compensate for the extra symmetries required to go from conformal supergravity to Poincar{\'e} supergravity. They also provide the necessary gauge fields which play the role of graviphotons in Poincar{\'e} supergravity. Typically these multiplets come with the wrong sign of the kinetic term in their action in conformal supergravity. This ensures that the Einstein-Hilbert term and its supersymmetrization come with the right sign of the kinetic term. There are also vector multiplets that play the role of physical matter multiplets coupled to $\cN=4$ Poincar\'e supergravity. Such vector multiplets always come with the right sign of the kinetic term in their action in conformal supergravity so that their kinetic terms remains with the right sign even in Poincar{\'e} supergravity.  

The Weyl multiplet is an off-shell multiplet. It contains superconformal gauge and matter fields that describe $128+128$ bosonic and fermionic off-shell degrees of freedom. The independent gauge fields include the vierbein $e_{\mu}^a$, the gravitini $\psi_{\mu}^i$ ($Q$-supersymmetry gauge fields), the $SU(4)$ gauge field ${{V_{\mu}}^i}_{j}$ and the dilatation gauge field $b_{\mu}$. The composite or dependent gauge fields are the spin connection ${\omega_{\mu}}^{ab}$, the gauge field associated with the special conformal transformation $f_{\mu}^{a}$, the chiral $U(1)$ gauge field $a_{\mu}$ and $S$-supersymmetry gauge field $\phi_{\mu}^{i}$, where $i,j=1,\dots,4$.
\begin{table}[]
\small
\centering
\begin{tabular}{ |p{1cm}| p{4cm} |p{5cm}| p{1 cm} |p{1cm} |p{1cm}|}
 \hline
  Fields& Symmetries(Generators)& Name/Restrictions& SU(4)& $w$ & $c$\\
 \hline
 ${e}_{a}^{\mu}$& Translations($P$)& vierbein& \textbf{1}& -1&0\\ \hline
 ${\omega_\mu}^{ab}$& Lorentz($M$)& spin connection& \textbf{1}& 0&0\\ \hline
 $b_{\mu}$&Dilatations& Dilatation gauge field& \textbf{1}&0&0\\ \hline
 ${{V_{\mu}}^i}_{j}$& SU(4)& SU(4) gauge field;\newline
 ${{V_{\mu}}^i}_{i}=0$,
 \newline
 ${V_{\mu i}}^{j}\equiv ({{V_{\mu}}^i}_{j})^*=-{{V_{\mu}}^j}_{i}$ &15&0&0\\ \hline
 ${f_{\mu}}^a$&conformal boosts& $K$-gauge field &\textbf{1}&1&0\\ \hline
 $a_{\mu}$& $U(1)$& $U(1)$ gauge field & \textbf{1}&0&0\\ \hline
 $\phi_{\alpha}$&&
 $\phi_{\alpha}\phi^{\alpha}=1, \phi^1=\phi^*_{1},$\newline
 $\phi^2=-\phi^*_{2}$&\textbf{1}&0&-1\\ \hline
 $E_{ij}$& &$E_{ij}=E_{ji}$& $\overline{\textbf{10}}$&1&-1\\ \hline
 ${T_{ab}}^{ij}$& &
 $T_{ab}{}^{ij}=-\frac{1}{2}
{\varepsilon_{ab}}^{cd}T_{cd}{}^{ij}$;\newline
 ${T_{ab}}^{ij}={T_{ab}}^{ji}$&\textbf{6} &1 &-1\\\hline
  ${D^{ij}}_{kl}$& & ${D^{ij}}_{kj}=0$\newline
 ${D^{ij}}_{kl}=\frac{1}{4}\varepsilon^{ijmn}\varepsilon_{klpq}{D^{pq}}_{mn}$\newline
 ${D}_{kl}^{ij}\equiv ({D^{kl}}_{ij})^*={D^{ij}}_{kl}$& $\textbf{20}'$&2 &0\\
 \hline
 $\phi_{\mu i}$ & $S$-supersymmetry & $S$-gauge field; $\gamma_{5}\phi_{\mu i}=\phi_{\mu i}$ & $\overline{\textbf{4}}$ & $1/2$ &$1/2$\\ \hline
 ${\psi_{\mu}}^{i}$& $Q$-supersymmetry & gravitini; $\gamma_5 {\psi_{\mu}}^{i}=\psi_{\mu}^{i}$ & $\textbf{4}$ & $-1/2$ & $-1/2$\\ \hline
 $\Lambda_{i}$& & $\gamma_{5}\Lambda_{i}=\Lambda_{i}$ &$\overline{\textbf{4}}$ & $1/2$ & $-3/2$\\ \hline
 ${\chi^{ij}}_{k}$ & & ${\chi^{ij}}_{j}=0$;\newline
$\gamma_{5}{\chi^{ij}}_{k}={\chi^{ij}}_{k}$, ${\chi^{ij}}_{k}=-{\chi^{ji}}_{k}$ & $\textbf{20}$ & $3/2$ & $-1/2$\\
\hline
\end{tabular}
\caption{Field Content of $\mathcal{N}=4$ Weyl Multiplet}
 \label{tab:weyl4}
\end{table}
Apart from the gauge fields, the Weyl multiplet contains several covariant matter fields: both bosonic as well as fermionic. Table \ref{tab:weyl4} summarizes all the fields of the Weyl multiplet along with their algebraic restrictions, $SU(4)$ representations, Weyl weight $w$ under local dilatations and $U(1)$ chiral weight $c$. The notations and conventions followed here are those of \cite{Butter:2019edc}.

The doublet of complex scalars $\phi_{\alpha}$ (where $1\leq\alpha\leq2$)  have rigid $SU(1,1)$ action and local chiral $U(1)$ action. They satisfy an $SU(1,1)\times U(1)$ invariant constraint $\phi_{\alpha}\phi^{\alpha}=1$, where $\phi^{\alpha}\equiv\eta^{\alpha\beta}(\phi_\beta)^*$ and $\eta=\text{diag}(1,-1)$. They parametrize an $SU(1,1)/U(1)$ coset manifold and hence are referred to as the coset scalars. Geometrically these coset scalars are described by the Maurer-Cartan 1-form $P_{\mu}$, the conjugate $\Bar{P}_{\mu}$ and the composite $U(1)$ gauge field $a_{\mu}$, defined below
\begin{align}\label{composite}
 P_{\mu} &= \varepsilon_{\alpha\beta} \phi^{\alpha} D_{\mu} \phi^{\beta} \notag\\
    \Bar{P}_{\mu} &= -\varepsilon^{\alpha\beta} \phi_{\alpha} D_{\mu} \phi_{\beta} \notag\\
    a_{\mu} &= i \phi^\alpha \partial_{\mu} \phi_{\alpha} +\frac{i}{4} \Bar{\Lambda}^i \gamma_{\mu} \Lambda_{i}
\end{align}
where $\varepsilon^{\alpha\beta}$ and $\varepsilon_{\alpha\beta}$ are the completely antisymmetric Levi-Civita in the $SU(1,1)$ space defined as $\varepsilon_{12}=\varepsilon^{12}=+1$. The $D_{\mu}$ appearing here and elsewhere  denotes the fully superconformal covariant derivative. The covariant derivative of the coset scalars satisfy the following constraint due to the composite nature of the $U(1)$ gauge field $a_{\mu}$.
\begin{align}
    \phi^{\alpha}D_{\mu}\phi_{\alpha}=-\frac{1}{4}\bar{\Lambda}^i\gamma_{\mu}\Lambda_i
\end{align}
The $Q$- and $S$- supersymmetry transformations, parametrized by $\epsilon^i$ and $\eta^i$ respectively, of the entire Weyl multiplet can be found in \cite{Bergshoeff:1980is,Ciceri:2015qpa,Butter:2019edc,Ciceri2025}. For our purpose, we need these transformations for the fermions, which we give below.
 \begin{align}\label{trans}
   \delta \psi_\mu{}^i=\,&2\,\mathscr{D}_\mu\epsilon^i
   -\tfrac12\gamma^{ab}T_{ab}{\!}^{ij}\gamma_\mu\epsilon_j+\varepsilon^{ijkl}\,
   \bar\psi_{\mu j} \epsilon_k\, \Lambda_l -\gamma_\mu\eta^i \,,\nonumber\\ 
 \delta \chi^{ij}{\!}_k=\,&-\tfrac12\gamma^{ab}
   \slashed{D}T_{ab}{\!}^{ij}\epsilon_k
   -\gamma^{ab} R(V)_{ab}{}^{[i}{}_k \,\epsilon^{j]}
   -\tfrac12\varepsilon^{ijlm}\,\slashed{D} E_{kl}\,\epsilon_m
   +D^{ij}{\!}_{kl}\, \epsilon^l\nonumber\\
   &-\tfrac1{6}\varepsilon_{klmn}E^{l[i}\gamma^{ab}\big[T_{ab}{\!}^{j]n}\epsilon^m+T_{ab}{\!}^{mn}\epsilon^{j]}\big]
   +\tfrac12 E_{kl}\,E^{l[i}\,\epsilon^{j]}
   -\tfrac12 \varepsilon^{ijlm}\bar{\slashed{P}}\gamma_{ab}
   T^{ab}{\!}_{kl}\,\epsilon_m\nonumber\\   
   &+\tfrac14 \gamma^a\epsilon_n \big[2\,\varepsilon^{ijln}\bar\chi^{m}{}_{lk}
   -\varepsilon^{ijlm}\bar\chi^{n}{}_{lk}\big]\gamma_a\Lambda_m
   +\tfrac14 \epsilon^{[i} \big[2\,\bar\Lambda^{j]}\slashed{D}\Lambda_k
   +\bar\Lambda_k\slashed{D}\Lambda^{j]}\big]\nonumber\\   
   &-\tfrac14\gamma^{ab}\epsilon^{[i}
   \big[2\, \bar\Lambda^{j]}\gamma_aD_{b}\Lambda_k
   - \bar\Lambda_k\gamma_a D_b\Lambda^{j]}\big]
   -\tfrac{5}{12}\varepsilon^{ijlm}\Lambda_m\,\bar\epsilon_l
   \big[E_{kn}\Lambda^n-2\,\slashed{P} \Lambda_k\big]\nonumber\\ 
   &+\tfrac{1}{12}\varepsilon^{ijlm} \Lambda_m\,\bar\epsilon_k\,
   \big[E_{ln}\Lambda^n  -2\,\slashed{P}\Lambda_l\big]
   -\tfrac12\gamma^{ab}T_{ab}{\!}^{ij} \gamma^c \epsilon_{[k}\,
   \bar\Lambda^l\gamma_c\Lambda_{l]} \nonumber\\
   &\,-\tfrac12\gamma^{ab} T_{ab}{\!}^{l[i} \gamma^c\epsilon_{[k}\,
   \bar\Lambda^{j]}\gamma_c\Lambda_{l]} 
    +\tfrac12\epsilon^{[i}\bar\Lambda^{j]}\Lambda^m\,
   \bar\Lambda_{k}\Lambda_m+\tfrac{1}{2} 
    T_{ab}{\!}^{ij}\,\gamma^{ab}\eta_k\,-\tfrac{1}{2}\varepsilon^{ijlm}E_{kl}\,\eta_m  \nonumber \\
    &\,-\tfrac{1}{4}\bar\Lambda_k\gamma^a\Lambda^{[i}\gamma_a\eta^{j]} -(\text{traces})\,,\nonumber\\
     \delta \Lambda_i=\,&-2\,\bar{\Slash{P}}\epsilon_i
   +E_{ij}\epsilon^j+\tfrac12\varepsilon_{ijkl}\,T_{bc}{\!}^{kl} 
   \gamma^{bc} \,\epsilon^j \,,
\end{align}
The supercovariant curvatures associated with the gauge fields $e_{\mu}{}^{a}$, $\omega_{\mu}{}^{ab}$ and $\psi_{\mu}{}^i$, can be read off from their gauge transformations and are given as follows:
\begin{subequations}
\begin{align}
    R(P)_{\mu\nu}{}^{a}=\,&2\mathscr{D}_{[\mu}e_{\nu]}{}^a-\Bar{\psi}_{[\mu}{}^i\gamma^a \psi_{\nu]i}, \label{eq:RP} \\
    R(M)_{\mu\nu}{}^{ab}=\,&2 \partial_{[\mu}\omega_{\nu]}{\!}^{ab}-2\omega_{[\mu}{\!}^{ab}\omega_{\nu] c}{\!}^b-4e_{[\mu}{\!}^{[a}f_{\nu]}{\!}^{b]} \nonumber \\
    &+\frac{1}{2}[\Bar{\psi}_{[\mu}\gamma^{ab}\phi_{\nu]i}-2\Bar{\psi_{[\mu i}}\gamma_{\nu]}R(Q)^{ab i}+ 2\Bar{\psi}_{\mu i}\psi_{\nu j}T^{ab ij}+\text{h.c}], \label{eq:RM} \\
    R(Q)_{\mu\nu}^{i}=\,& 2\mathscr{D}_{[\mu}\psi_{\nu]}^i-\gamma_{[\mu}\phi_{\mu]}{\!}^i-\frac{1}{2}\gamma^{ab}\gamma_{[\mu}\psi_{\nu]j}T_{ab}{\!}^{ij}+\frac{1}{2}\varepsilon^{ijkl}\Bar{\psi}_{[\mu j}\psi_{\nu]k}\Lambda_l\;, \label{eq:RQ}
\end{align}
\end{subequations}
where, the derivative $\mathscr{D}_{\mu}$ is covariant with respect to all bosonic symmetries except the special conformal transformation. The curvatures satisfy the following constraints:
\begin{subequations}\label{eq:Constraints}
\begin{align}
    R(P)_{\mu\nu}{}^{a} &= 0, \label{eq:RPZero} \\
    e^{\nu}{}_{b}R(M)_{\mu\nu}{}^{ab} &= 0, \label{eq:RMZero} \\
    \gamma^{\mu}R(Q)_{\mu\nu}^{i} &= 0. \label{eq:RQZero}
\end{align}
\end{subequations}

The above constraints make some of the gauge fields such as $\omega_{\mu}^{ab}$, $f_{\mu}^a$ and $\phi_{\mu}^i$ dependent and their expressions can be found by solving the above constraints \eqref{eq:Constraints}. In particular the expression for the gauge field $f_{\mu}^a$ corresponding to the special conformal transformation can be obtained by solving the constraint \eqref{eq:RMZero} and is given as below (up to fermionic terms which we have omitted).
\begin{align}\label{eq:FM}
    f_{\mu}{\!}^a &= \frac{1}{2} R_{\mu}{\!}^{a}-\frac{1}{12}Re_{\mu}{\!}^{a}\;. 
\end{align}
In the above equations, we have used the following definitions:
\begin{align}
   R_{\mu}{}^{a}=e^{\nu}{}_{b}R_{\mu\nu}{}^{ab}\;, \;\;  R=e^{\mu}{}_{a}R_{\mu}{}^{a}
\end{align}
where $R_{\mu\nu}{}^{ab}\equiv \left(R(M)_{\mu\nu}{}^{ab}\right)_{|f=0}$. In the Poincar{\'e} gauge $b_\mu=0$, $R_{\mu\nu}{}^{ab}$, $R_{\mu}{}^{a}$ and $R$ have the interpretation of Riemann tensor, Ricci tensor and Ricci scalar respectively. The supersymmetry transformations (both $Q$ as well as $S$) of all the curvatures can be obtained and are given in \cite{Bergshoeff:1980is,Ciceri:2015qpa,Butter:2019edc}. For our purpose, we need the transformation for the fermionic curvature $R(Q)_{ab}^i$ which we give below.
\begin{align}\label{rqvary}
    \delta {R(Q)_{ab}}^{i}=\,&-\frac{1}{2}R(M)_{abcd}\gamma^{cd}\epsilon^{i}+\frac{1}{4}(\gamma^{cd}\gamma_{ab}+\frac{1}{3}\gamma_{ab}\gamma^{cd})[R(V)_{cd}{\!}^i{\!}_{j}\epsilon^j+\tfrac12 i F_{cd}\epsilon^i+\slashed{D}T_{cd}{\!}^{ij}\epsilon_{j}]&\nonumber\\&+\frac{1}{2}(\gamma^{cd}\gamma_{ab}+\tfrac13 \gamma_{ab}\gamma^{cd})T_{cd}{\!}^{ij}\eta_{j}
\end{align}    
The chiral $U(1)$ field strength $F_{ab}$, appearing in the above transformation rules, is given by 
\begin{equation}\label{fab}
    F_{ab}=2i \Bar{P}_{[a}P_{b]}-\tfrac12 i[\Bar{\Lambda}^i\gamma_{[a}D_{b]}\Lambda_i-\text{h.c}]\;.
\end{equation}
For future reference, we also give the supersymmetry transformation of the dependent gauge field $\phi_{\mu}^{i}$:
\begin{align}\label{transphimu}
\delta\phi_{\mu}^{i}=\,&2\mathscr{D}_{\mu}\eta^{i}-\frac{1}{6}\gamma_{\mu}\gamma^{ab}T_{ab}{}^{ij}\eta_{j}-2f_{\mu}{}^{a}\gamma_{a}\epsilon^{i}+\frac{1}{4}T_{ab}{}^{ij}T^{cd}{}_{jk}\gamma_{cd}\gamma_{\mu}\gamma^{ab}\epsilon^{k}\nonumber\\
    \,&+\frac{1}{6}\Big(\gamma_{\mu}\gamma^{ab}-3\gamma^{ab}\gamma_{\mu}\Big)\Big(R(V)_{ab}{}^{i}{}_{j}\epsilon^{j}+\frac{1}{2}iF_{ab}\epsilon^{i}+\frac{1}{2}D_{a}T_{cd}{}^{ij}\gamma_{b}\epsilon_{j}\Big)+\cdots
\end{align}
where the ellipsis contains fermionic terms which are not relevant for us.

$\mathcal{N}=4$ vector multiplet is an on-shell multiplet \cite{deRoo:1984zyh}. It contains a gauge field $A_{\mu}$, scalar fields $\phi_{ij}$, and spin-1/2 fermion $\psi_{i}$ (gaugino). Their properties are summarized in Table \ref{tab:vector}.
\begin{table}[H]
\small
    \centering
    \begin{tabular}{|c|c|c|c|c|
}
    \hline
         Field& Properties& $SU(4)$ &$w$ &$c$ \\
         \hline
         $A_{\mu}$ & Gauge field & \textbf{1} &0&0\\ \hline
         $\psi_i$& $\gamma_5 \psi_i=-\psi_i$ & $\overline{\textbf{4}}$ & $3/2$& $-1/2$\\  \hline
         $\phi_{ij}$& $\phi^{ij}\equiv(\phi_{ij})^*=-\tfrac12 \varepsilon^{ijkl}\phi_{kl}$& \textbf{6} & 1 &0\\
         \hline
    \end{tabular}
    \caption{$\mathcal{N}=4$ Vector multiplet}
    \label{tab:vector}
\end{table}
The $Q$- and $S$-supersymmetry transformations for $n$ abelian vector multiplets labeled by $I=1,2,\cdots,n$ are given by
\begin{align}\label{transvec}
    \delta A_{\mu}^I=\,& \Phi[\Bar{\epsilon}^i \gamma_{\mu}\psi_i^I-2\Bar{\epsilon}^i\psi_{\mu}{\!}^j\phi_{ij}^{I}+\Bar{\epsilon}_{i}\gamma_{\mu}\Lambda_j\phi^{I ij}]+\text{h.c.}\,,\nonumber\\
    \delta \psi_{i}^{I}=\,&-\frac{1}{2\Phi}\gamma^{ab}\epsilon_{i}\Big(\hat{F}_{ab}^{I +}-\frac{\Phi^*}{4}\Bar{\Lambda}^{i}\gamma_{ab}\psi_{i}+\Phi^*T_{ab ij}\phi^{I ij}\Big)-2\slashed{D}\phi_{ij}^{I}\epsilon^j+\epsilon_{k}E_{ij}\phi^{I jk}&\nonumber\\& +\frac{1}{2}\epsilon_{i}\Bar{\Lambda}_{j}\psi^{I j}-\epsilon_{j}\Bar{\Lambda}_{i}\psi^j+\frac{1}{2}\gamma^a\epsilon^j\Bar{\Lambda}^{k}\gamma_a\Lambda_{i}\phi_{jk}^{I}-2\phi_{ij}^{I}\eta^{j}\,,\nonumber\\
    \delta \phi_{ij}^{I}=\,& 2\Bar{\epsilon}_{[i}\psi_{j]}^{I}-\varepsilon_{ijkl}\Bar{\epsilon}^k\psi^{I l}\;,
    \end{align}
    where,
    \begin{equation}
        \Phi=\phi^1+\phi^2, ~~~\Phi^*=\phi_1-\phi_2\;,
    \end{equation}
and $\hat{F}_{\mu\nu}^{I}$ is the supercovariant field strength associated with the gauge fields $A_{\mu}^I$ and is given as:
    \begin{equation}
\hat{F}_{\mu\nu}^{I}=\partial_{[\mu}A_{\nu]}^{I}-\Phi\Big(\Bar{\psi}_{[\mu i}\gamma_{\nu]}\Lambda_{j}\phi^{I ij}+\Bar{\psi}^{i}_{[\mu}\gamma_{\nu]}\psi_{i}^{I}-4\Bar{\psi}^{i}_{\mu}\psi^{j}_{\nu}\phi^{I}_{ij}\Big)+\text{h.c}\,,
    \end{equation}
 We now provide a very brief outline of the construction of $\cN=4$  higher derivative Poincar\'e supergravity within the framework of conformal supergravity constructed using the standard Weyl multiplet. Such theories contains a class of higher derivative terms that contains all terms related to Weyl square terms by supersymmetry:
\begin{itemize}
    \item One starts with the Lagrangian 
    \begin{align}
        \cL=\cL_{V}+ \lambda \cL_{CSG}
    \end{align}
    where $\cL_{V}$ is the Lagrangian for $(6+n_V)$ vector multiplets coupled to conformal supergravity \cite{deRoo:1984zyh, Ciceri2025} and $\cL_{CSG}$ is the pure conformal supergravity invariant \cite{Butter:2016mtk, Butter:2019edc}.
    \item One breaks the extra symmetries (Special conformal transformation, dilatation and $S$-supersymmetry) by imposing the following gauge fixing conditions \cite{deRoo:1984zyh, Ciceri2025}:
    \begin{table}[H]
\small
    \centering
    \begin{tabular}{|c|c|c|
}
    \hline
         Symmetries& Gauge fixing conditions &Name \\
         \hline
       SCT& $b_{\mu}=0$& $K$-gauge\\ \hline
        Dilatation & $\phi^{Iij}\phi_{ij}^J\eta_{IJ}=-\frac{6}{\kappa^2}$ &$D$-gauge\\ \hline
        $S$-supersymmetry & $\phi^{Iij}\psi^{J}_{j}\eta_{IJ}=0$  &$S$-gauge\\
         \hline
    \end{tabular}
    \caption{Gauge fixing conditions}
    \label{tab:gaugecon}
\end{table}
    $\kappa=\sqrt{8\pi G}$ , where $G$ is the Newton's constant.
    \item  One eliminates the ``auxiliary fields'' ($T_{ab}{}^{ij}, E_{ij}, D^{ij}{}_{kl}, \chi^{ij}{}_{k}$) by using their equations of motion. This generates a derivative expansion of the action. For example, eliminating the auxiliary fields by their leading-order equations of motion, generates four derivative corrections to $\cN=4$ Poincar\'e supergravity. All the auxiliary field equations and four derivative corrections to $\cN=4$ Poincar\'e supergravity can be found in \cite{Ciceri2025}. For our purpose, we need the $D^{ij}{}_{kl}$ field equations. This field equation combined with the dilatation gauge fixing condition gives the following
\begin{equation}\label{dijkl}
    \phi^{I ij}\phi^{J}_{ 
 kl}\eta_{IJ}= -\frac{1}{\kappa^2}\delta^{i}_{[k}\delta^{j}_{l]}+ \lambda X^{ij}{}_{kl}
\end{equation}
where 
\begin{equation}
    X^{ij}{}_{kl}\equiv -\mathcal{H}D^{ij}{}_{kl}+ \frac{1}{2}\mathcal{D}\mathcal{H}\Big[T^{abmn}T_{ab}{}^{ij}\varepsilon_{klmn}-\frac{1}{2}E_{km}E_{ln}\varepsilon^{ijmn} \Big]_{\mathbf{20'}}+\text{h.c.}
\end{equation}
Here $\cH(\phi_\alpha)$ is a holomorphic function of the coset scalars and homogeneous function of degree zero. The derivative operator acting on $\cH$ is defined as 
\begin{equation}
    \cD\mathcal{H}=-\varepsilon_{\alpha\beta}\phi^\alpha \frac{\partial\mathcal{H}}{\partial \phi_{\beta}}\;.
\end{equation}
Ideally one should use a gauge fixing condition on the vector multiplet scalars $\phi^{I}_{ij}$ to break $SU(4)$ $R$-symmetry in order to go from conformal supergravity to Poincar{\'e} supergravity. Alternatively, one can work with a combination of the scalar fields that is invariant under the $SU(4)$ $R$-symmetry defined as follows:
\begin{align}\label{Mdef}
        M^{IJ}=\frac{1}{4}\phi^{I ij}\phi_{ij}^{J},
    \end{align} 
It turns out that the bosonic part of the Poincar{\'e} supergravity action can be written completely in terms of the above-mentioned matrix valued scalar fields. If one takes the leading order $\cO(\lambda^0 )$ values of the scalar fields, then the $SU(4)$ invariant matrix valued scalar field satisfies the properties given below.
    \begin{align}\label{Mprop}
        M^{IK}\eta_{KL}M^{LJ}=- \,&\frac{1}{4}M^{IJ}\\ 
        Tr(M\eta)=\,&-\frac{3}{2}
    \end{align}
\end{itemize}
One can also check that, with the above mentioned properties, the matrix valued scalar field is parametrized by $6n_v$ physical scalar fields.
\section{\texorpdfstring{$\mathcal{N}=4$ Fully Supersymmetric Solution}{N=4 Fully Supersymmetric Solution}}
\label{N4susy}
In this section, we find the fully supersymmetric solution in $\cN=4$ higher derivative Poincar\'e supergravity that is considered in the previous section. A fully supersymmetric $\cN=4$ solution implies that the background geometry
possesses four covariantly constant Killing spinors ($16$ conserved supercharges). We assume the existence of such spinors and demand the $Q$- supersymmetry transformation of the fermions to vanish. The vanishing $Q$-supersymmetry variations of the fermions put stringent conditions on the bosonic fields. We have to find field configurations that satisfy these conditions.

Since we are in the superconformal formalism, the fermions also transform under $S$-supersymmetry. We can only demand the fermions to vanish under $Q$-supersymmetry up to an $S$- transformation. An alternative way to surpass this is to find combinations of spinors that are $S$-invariant. This is done by first identifying a spinor that transforms as a pure gauge under $S$-supersymmetry. A linear combination of other spinors with this spinor can give us $S$-invariant spinors. We construct such $S$-invariant spinors and demand that their $Q$-supersymmetry transformations vanish. This procedure is independent of the $S$-gauge. We are going to follow this procedure along the lines of \cite{Mohaupt:2000mj, LopesCardoso:2000qm}.

The spinor that transforms as a pure gauge under $S$-supersymmetry can be obtained using the fields of $\mathcal{N}=4$ vector multiplet as shown below:
\begin{equation}\label{in}
    \zeta_{i}= 2\phi^{-2} \phi_{ij}^{I}{\psi^{j J}}\eta_{IJ},~\phi^2=\phi^{Iij}\phi^{J}_{ij}\eta_{IJ}
\end{equation}
This spinor transforms under $Q$- and $S$- supersymmetry as 
\begin{flalign}\label{zeta}
    \delta \zeta^{i}= 2\phi^{-2}\phi^{I ij}\left[-\frac{1}{2\Phi}\gamma^{ab}\epsilon_{j}\left(\hat{F}_{ab}^{+J}+\Phi^{*}T_{ablk}\phi^{Jlk}\right)-2\epsilon^{k}\slashed D\phi^{J}_{jk} + \epsilon_{k}E_{jl}\phi^{Jlk}\right]\eta_{IJ}+\eta^{i}.
\end{flalign}

We now start analyzing the fermionic variations. We begin with the Weyl multiplet fermion $\Lambda_{i}$. Since it does not transform under $S$-supersymmetry, we do  not need to add any $S$-supersymmetry compensator. The equation that we would obtain by setting its $Q$-supersymmetry transformations to zero is the following.
\begin{align}\label{lambdavari}
    \delta \Lambda_i= -2\Bar{\slashed P}\epsilon_{i}+ E_{ij}\epsilon^j+\frac{1}{2}  \varepsilon_{ijkl}{T_{bc}}^{kl}\gamma^{bc}\epsilon^{j}\overset{!}{=}0
    \end{align}
We now collect terms according to their  $SU(4)$ and Lorentz representations and independently set them to zero. By doing so, we obtain the following equations on the bosonic fields.
\begin{subequations}\label{con}
\begin{align} 
\Bar{P}_{\mu}=-\varepsilon^{\alpha\beta}\phi_{\alpha}D_{\mu}\phi_{\beta}=\,&0,\label{con1} \\ 
    E_{ij}=\, &0,\label{con2}\\ 
    T_{ab}{\!}^{ij}=\,& 0\label{con3}
\end{align}
\end{subequations}
In the absence of fermions, the condition \eqref{con1} becomes
\begin{equation}
\varepsilon^{\alpha\beta}\phi_{\alpha}\partial_{\mu}\phi_{\beta}=0,
\end{equation}
which implies 
\begin{align}\label{con4}
    \partial_{\mu}\phi_{\beta}=\lambda_{\mu}\phi_{\beta}.
\end{align}
\eqref{con4} has a general solution
\begin{equation}
    \phi_{\beta}(x)=\phi_{\beta}^{0}e^{\lambda_{\mu}x^{\mu}}
\end{equation}
where $\phi_{\beta}^{0}$ is a constant. The above solution is gauge equivalent to the constant $\phi_{\beta}^{0}$. Hence, we can choose our coset scalars to be constant as a solution to \eqref{con1}.

Next we turn to the variation of $\chi^{ij}{\!}_{k}$. The bosonic terms in the $S$-transformation of $\chi^{ij}{}_{k}$ \eqref{trans} vanishes upon using \eqref{con}. Hence, one does not need any explicit contributions coming from the $S$-supersymmetry compensator $\zeta^i$. The $Q$-supersymmetry of $\chi^{ij}{}_{k}$ also simplifies considerably upon using \eqref{con}. As a consequence, we obtain: 
\begin{equation}
     \delta {\chi^{ij}}_{k}= -\gamma^{ab}{{R(V)_{ab}}^{[i}}_{k}\epsilon^{j]}+\frac{1}{4}\delta^{j}_{k}\gamma^{ab}R(V)_{ab}{}^{i}{}_{j}\epsilon^j+{D^{ij}}_{kl}\epsilon^{l}\overset{!}{=}0,
\end{equation}
which implies
\begin{subequations}
    \begin{align}
{R(V)_{ab}{\!}^{i}}_{j}=\,& 0,\\
D^{ij}{\!}_{kl}=\,& 0.\label{DIJKL}
\end{align}
\end{subequations}

Now we turn to the $\psi_{\mu}^{i}$ (gravitini) variations. Since they are gauge fields, we demand the corresponding gauge invariant quantity, the field strengths $R(Q)_{\mu\nu}^{i}$, to vanish under supersymmetric variations. One can also see that the bosonic terms in the $S$-supersymmetry transformations of $R(Q)^i_{\mu\nu}$ vanishes upon using \eqref{con}. Hence one does not need the $S$-supersymmetry compensator. The $Q$-supersymmetry variation also simplifies upon using \eqref{con} and the expression for the $U(1)$ field strength \eqref{fab}. As a consequence, one obtains the following condition:
\begin{align}\label{rq}
    \delta {R(Q)_{ab}}^{i}=\,&-\frac{1}{2}R(M)_{abcd}\gamma^{cd}\epsilon^{i}\overset{!}{=}0
\end{align}
which implies that
\begin{equation}\label{rm0}
    R(M)_{\mu\nu}{\!}^{cd}=0.
\end{equation}

We now take the variation of the vector multiplet fermion $\psi_{i}^{I}$ (gaugino). The $Q$ supersymmetry transformation of $\psi_{i}^{I}$ simplifies using \eqref{con}, but the $S$-supersymmetry transformation does not vanish. Hence, we need to add the $S$-supersymmetry compensator $\zeta^i$. One can see that, the following combination
\begin{align}
    (\psi_{i}^{I}+ 2 \phi_{ij}^{I}\zeta^j)\equiv \vartheta^{I}_{i}
\end{align}
is $S$-invariant. Demanding the vanishing of the $Q$-supersymmetry variation of $\vartheta_{i}^{I}$ gives us the following conditions:
\begin{subequations}
    \begin{align}
\hat{F}_{ab}^{+I}=\,& 4\phi^{-2}M^{IK}\hat{F}_{ab}^{+L}\eta_{KL}=-\tfrac23 M^{IK}\hat{F}_{ab}^{+L}\eta_{KL}\;,\label{electriccon}\\
D_{\mu}\phi_{iq}^{I}= 4\,&\phi^{-2}M^{IK}D_{\mu} \phi_{iq}^{L}\eta_{KL}=-\tfrac23 M^{IK}D_{\mu} \phi_{iq}^{L}\eta_{KL}\;, \label{phicon}
\end{align}
\end{subequations}
where, in the second step of both the equations we have used the dilatation gauge fixing condition $\phi^2=-6$ in the units $\kappa=1$, which we will follow from here onward.  

We note that, the correction to the $D^{ij}{}_{kl}$ equations of motion  \eqref{dijkl} vanishes upon using the constraints on the bosonic field configurations obtained in \eqref{con2}, \eqref{con3} and \eqref{DIJKL}.
 \begin{align}
     X^{ij}{}_{kl}=0
 \end{align}
 As a consequence, $\phi_{ij}^{I}$ satisfies the leading order equations of motion and the matrix valued scalar field $M^{IJ}$ satisfies the property \eqref{Mprop}. This serves as a crucial step in our analysis and also implies that the final outcome of our analysis holds with or without the presence of higher derivative corrections. Contracting \eqref{electriccon} with an $M\eta$ and using the property \eqref{Mprop}, we get
\begin{align}
    M^{IK}\hat{F}^{+ L}_{ab}\eta_{KL}=0\;.
\end{align}
Using this back in \eqref{electriccon} implies 
\begin{align}\label{fab0}
    \hat{F}_{ab}^{+I}=0
\end{align}
Similarly contracting \eqref{phicon} with an $M\eta$ and using the property \eqref{Mprop}, we get
\begin{equation}\label{deriPhi}
D_{\mu}\phi_{ij}^{I}=0\;.
\end{equation}
 Since $\phi^{I}_{ij}$ is inert under special conformal transformation, the fully supercovariant derivative $D_{\mu}$ is the same as the covariant derivative $\mathscr{D}
 _{\mu}$ without the $K$-connection. Hence \eqref{deriPhi} also implies $\mathscr{D}_{\mu}\phi^{I}_{ij}=0$. The condition on the matrix valued scalar field $M^{IJ}$ \eqref{Mdef} that follows from this is
\begin{equation}\label{Mder}
\partial_{\mu}M^{IJ}=\mathscr{D}_{\mu}M^{IJ}=\phi^{I ij}\mathscr{D}_{\mu}\phi_{ij}^{I}+\phi_{ij}^{I}\mathscr{D}_{\mu}\phi^{I ij}=0\;.
\end{equation}
In the second step of the above equation, we have used the fact that the only non-trivial bosonic superconformal transformation possessed by $M^{IJ}$ is dilatation and because of the $K$-gauge condition $b_\mu=0$, we have $\partial_{\mu}M^{IJ}=\mathscr{D}_{\mu}M^{IJ}$. The equation \eqref{Mder} further implies: 
\begin{align}\label{mijcon}
    M^{IJ}=\text{constants}
\end{align}
Now we take the variation of the supercovariant derivative of the gaugino, $\psi_{i}^{I}$. This gives rise to a new condition. We can read off the fully supercovariant derivative of $\psi_{i}^{I}$ from the transformation rule given in \eqref{transvec}:
\begin{align}\label{deripsi}
D_{\mu}\psi_{i}^{I}=\,&\mathscr{D}_{\mu}\psi_{i}^{I}+\frac{1}{4\Phi}\gamma^{ab}\psi_{\mu i}\Big(\hat{F}^{+I}_{ab}+\Phi^{*}T_{ab kl}\phi^{I kl}\Big)+\slashed{D}\phi_{ij}^{I}\psi_{\mu}^{j}+\phi_{ij}^{I}\phi_{\mu}^{j}
\end{align}
The $S$ -supersymmetry transformation of \eqref{deripsi} is given by:
\begin{align}
    \delta_{S}(D_{\mu}\psi_{i}^{I})=\,&-2{D}_{\mu}\phi_{ij}^{I}\eta^{j}-\frac{1}{4\Phi}\gamma^{ab}\gamma_{\mu}\Big(\hat{F}^{+I}_{ab}+\Phi^{*}T_{ab kl}\phi^{I kl}\Big)\eta_{i}-\slashed{D}\phi_{ij}^{I}\gamma_{\mu}\eta^{j}\nonumber\\
    \,&-\frac{1}{6}\gamma_{\mu}\gamma^{ab}T_{ab}{}^{jk}\eta_{k}\phi_{ij}^{I}
\end{align}
Using the conditions we got from the previous analysis, one can readily see that, on the fully supersymmetric solution, 
\begin{align}
      \delta_{S}(D_{\mu}\psi_{i}^{I})=0
\end{align}
So, we do not have to add any $S$-supersymmetry compensator to \eqref{deripsi}. It is sufficient to take the $Q$ -supersymmetry variation of \eqref{deripsi}:
\begin{align}
    \delta_{Q}(D_{\mu}\psi_{i}^{I})=\,&\frac{1}{2\Phi^2}\gamma^{ab}{D}_{\mu}\Phi\Big(\hat{F}^{+I}_{ab}+\Phi^{*}T_{ab kl}\phi^{I kl}\Big)\epsilon_{i}-\frac{1}{2\Phi}\gamma^{ab}\Big[{D}_{\mu}\hat{F}^{+I}_{ab}+{D}_{\mu}(\Phi^{*}T_{ab kl}\phi^{I kl})\Big]\epsilon_{i}\nonumber\\
\,&-2{D}_{\mu}{D}_{\nu}\phi^{I}_{ij}\gamma^{\nu}\epsilon^{j}+D_{\mu}(E_{ij}\phi^{Ijk})\epsilon_{k}-\frac{1}{2}\slashed{D}\phi_{ij}^{I}\gamma^{ab}T_{ab}{}^{jk}\gamma_{\mu}\epsilon_{k}\nonumber\\
\,&-\frac{1}{8\Phi}\gamma^{ab}\gamma^{cd}T_{cd ij}\gamma_{\mu}\Big(\hat{F}_{ab}^{I}+\Phi^{*}T_{ab kl}\phi^{I kl}\Big)\epsilon^{j}\nonumber\\
\,&+\phi_{ij}^{I}\bigg[\frac{1}{4}T_{ab}{}^{jk}T^{cd}{}_{kl}\gamma_{cd}\gamma_{\mu}\gamma^{ab}\epsilon^{l}\nonumber\\
\,&+\frac{1}{6}\Big(\gamma_{\mu}\gamma^{ab}-3\gamma^{ab}\gamma_{\mu}\Big)\Big(R(V)_{ab}{}^{j}{}_{k}\epsilon^{k}+\frac{1}{2}iF_{ab}\epsilon^{k}+\frac{1}{2}D_{a}T_{cd}{}^{jk}\gamma_{b}\epsilon_{k}\Big)
\bigg]\overset{!}{=}0
\end{align}
Again using the previous results, the only non-trivial condition we get, is 
\begin{align}\label{ddphi}
   {D}_{\mu}{D}_{\nu}\phi^{I}_{ij}=0
\end{align}
Combining the above with \eqref{deriPhi}, we obtain
\begin{align}\label{fmu0}
    f_{\mu}{}^{a}=0
\end{align}
This in conjunction with the condition \eqref{rm0}, implies that all the components of the Riemann curvature tensor vanish i.e.,
\begin{align}
        R(M)_{\mu\nu}{\!}^{cd}=2 \partial_{[\mu}\omega_{\nu]}{\!}^{ab}-2\omega_{[\mu}{\!}^{ab}\omega_{\nu] c}{\!}^b= R_{\mu\nu}{}^{ab}=0
\end{align}
Based on our analysis in this section, we can conclude that there is a unique solution in $\cN=4$ supergravity with or without higher derivative corrections that is fully supersymmetric. In this solution, the underlying geometry is a flat spacetime with no electromagnetic fluxes. The coset scalars and the $6n_v$ vector multiplet scalars take arbitrary constant values in this background. We summarize the result in Table \ref{tab:my_label}. 
\begin{table}[H]
\small
    \centering
    \begin{tabular}{|c|c|}
    \hline
         Field& Full Supersymmetric solution\\
         \hline
         $F_{\mu\nu}^{I}$ & 0~(No flux)\\
         \hline
         $g_{\mu\nu}$&Flat spacetime\\
         \hline
         $M^{IJ}$ & constant\\
         \hline
         $\phi_{\alpha}$&constant\\
         \hline
    \end{tabular}
    \caption{$\mathcal{N}=4$ Fully supersymmetric solution}
    \label{tab:my_label}
\end{table}

\section{\texorpdfstring{$\mathcal{N}=3$ Higher Derivative Supergravity}{N=3 Higher Derivative Supergravity}}\label{N3sugra}
\begin{table}[H]
\small
\centering
\begin{tabular}{|p{1cm}|p{4cm}| p{4cm}| p{1 cm}| p{1cm}| p{1cm}|}
 \hline
 Fields& Symmetries(Generators)& Name/Restrictions& SU(3)& $w$ & $c$\\
 \hline
${e}_{a}^{\mu}$& Translations($P$)& vierbein& \textbf{1}& -1&0\\ \hline
 ${\omega_\mu}^{ab}$& Lorentz($M$)& spin connection& \textbf{1}& 0&0\\ \hline
$b_{\mu}$&Dilatations& Dilatation gauge field& \textbf{1}&0&0\\ \hline
${{V_{\mu}}^i}_{j}$& $SU(3)$& $SU(3)_{R}$ gauge field;\newline
 ${{V_{\mu}}^i}_{i}=0$,
 \newline
 ${V_{\mu}i}^{j}\equiv ({{V_{\mu}}^i}_{j})^*=-{{V_{\mu}}^j}_{i}$ &\textbf{8}&0&0\\ \hline
 ${f_{\mu}}^a$&conformal boosts& $K$-gauge field &\textbf{1}&1&0\\ \hline
$A_{\mu}$& $U(1)_R$& $U(1)_R$ gauge field & \textbf{1}&0&0\\ \hline
 $E_{i}$& & Complex & $\overline{\textbf{3}}$&1&-1\\ \hline
 ${T_{ab}}^{i}$& &
 $T_{ab}^{i}=\frac{1}{2}
{\varepsilon_{ab}}^{cd}T_{cd}^{i}$&\textbf{3} &1 &1\\ \hline
 ${D^{i}}_{j}$& &
 ${D}_{i}{\!}^{j}\equiv (D^{i}{\!}_{j})^*=D^{j}{\!}_{i}$& $\textbf{8}$&2 &0\\
 \hline
 $\phi_{\mu i}$ & $S$-supersymmetry & $S$-gauge field; $\gamma_{5}\phi_{\mu i}=\phi_{\mu i}$ & $\overline{\textbf{3}}$ & $1/2$ &$1/2$\\ \hline
 ${\psi_{\mu}}^{i}$& $Q$-supersymmetry & gravitini; $\gamma_5 {\psi_{\mu}}^{i}=\psi_{\mu}^{i}$ & $\textbf{3}$ & $-1/2$ & $-1/2$\\ \hline
$\Lambda_{L}$& & $\gamma_{5}\Lambda_{L}=\Lambda_{L}$ &$\textbf{1}$ & $1/2$ & $-3/2$\\ \hline
$\chi_{ij}$ & &$\gamma_5\chi_{ij}=\chi_{ij}$ & $\overline{\textbf{6}}$ & $3/2$ & $-1/2$\\ \hline
$\zeta^i$& &$\gamma_{5}\zeta^{i}=\zeta^{i}$&\textbf{3} &$3/2$& $-1/2$\\
\hline
\end{tabular}
\caption{Field Content of $\mathcal{N}=3$ Weyl Multiplet}
\label{tab:weyl3}
\end{table}

In this section, we review four dimensional $\mathcal{N}=3$ supergravity using the framework of conformal supergravity. For a detailed construction of $\cN=3$ conformal supergravity, see \cite{vanMuiden:2017qsh,Hegde:2018mxv, Hegde:2021rte}. Using these results, a class of higher derivative $\mathcal{N}=3$ Poincar\'e supergravity was constructed in \cite{Hegde:2022wnb}, that contains all the terms related to Weyl squared term by supersymmetry. The crucial ingredients are the $\mathcal{N}=3$ Weyl and vector multiplets. A Poincar\'e supergravity coupled to $n_V$ vector multiplets is realised as a Weyl multiplet coupled to $n_V+3$ vector multiplets in conformal supergravity where the three compensating vector multiplets have a wrong sign kinetic term. As explained in the introduction, one gauge fixes additional symmetries using the compensators and solves the auxiliary field equations to write the action in Poincar\'e supergravity variables. In \cite{Hegde:2022wnb} these further steps were carried out only for the case of pure supergravity. However, for our analysis it is not necessary to eliminate the auxiliary fields since our supersymmetry analysis for the solution is best done in off-shell conformal supergravity variables. We will proceed therefore by considering fully supersymmetric solutions corresponding to the action in conformal supergravity with $n_V+3$ vector multiplets, which is equivalent to higher derivative matter coupled Poincar\'e supergravity.

 The $\mathcal{N}=3$ Weyl multiplet is an off-shell multiplet and contains superconformal gauge and matter fields that describe $64+64$ bosonic and fermionic off-shell degrees of freedom. The independent gauge fields include the vierbein $e_{\mu}^a$, the gravitini $\psi_{\mu}^i$ ($Q$-supersymmetry gauge fields), the $SU(3)$ gauge field ${{V_{\mu}}^i}_{j}$ and the dilatation gauge field $b_{\mu}$. The composite or dependent gauge fields are the spin connection ${\omega_{\mu}}^{ab}$, the gauge field associated with the special conformal transformation $f_{\mu}^{a}$, the chiral $U(1)$ gauge field $A_{\mu}$ and $S$-supersymmetry gauge field $\phi_{\mu}^{i}$, where $i,j=1,2,3$. Apart from the gauge fields, the Weyl multiplet contains several covariant matter fields: both bosonic as well as fermionic. Table \ref{tab:weyl3} summarizes all the fields of the Weyl multiplet along with their algebraic restrictions, $SU(3)$ representations, Weyl weight $w$ under local dilatations and $U(1)$ chiral weight $c$. We follow the notations and conventions of \cite{Hegde:2022wnb}.

The $Q$- and $S$- supersymmetry transformations, parametrized by $\epsilon^i$ and $\eta^i$ respectively, of the entire Weyl multiplet can be found in \cite{Hegde:2018mxv, Hegde:2021rte, Hegde:2022wnb}. For our purpose, we need these transformations for the fermions, which we give below. 
\begin{align}\label{N=3trans}
    \delta\psi_{\mu}{}^{i}=\,&2\mathscr{D}_{\mu}\epsilon^i-\frac{1}{8}\varepsilon^{ijk}\gamma.T_{j}\gamma_{\mu}\epsilon_{k}-\varepsilon^{ijk}\Bar{\epsilon}_{j}\psi_{\mu k}\Lambda_{L}-\gamma_{\mu}\eta^{i}\,\nonumber\\
    \delta\Lambda_L=\,& -\frac{1}{4}E_{i}\epsilon^i+\frac{1}{4}\gamma.T_i\epsilon^i\,\nonumber\\
    \delta \chi_{ij}=\,&2\slashed{D}E_{(i}\epsilon_{j)}-8\varepsilon_{kl(i}\gamma.R(V)^l{}_{j)}\epsilon^k-2\gamma.\slashed{D}T_{(i}\epsilon_{j)}+\frac{1}{3}\varepsilon_{kl(i}D^l{}_{j)}\epsilon^k\,\nonumber\\
    &+\frac{1}{4}\varepsilon_{kl(i}E^k\gamma\cdot T_{j)}\epsilon^l-\frac{1}{3}\Bar{\Lambda}_{L}\gamma_{a}\epsilon_{(i}\gamma^a\zeta_{j)}+\frac{1}{4}\varepsilon_{lm(i}E_{j)}E^m\epsilon^l-\Bar{\Lambda}_{L}\gamma^a\Lambda_{R}\gamma_{a}E_{(i}\epsilon_{j)},\nonumber\\
    &-\Bar{\Lambda}_{L}\gamma\cdot T_{(i}\gamma^a\Lambda_{R}\gamma_a\epsilon_{j)}+2 \gamma\cdot T_{(i}\eta_{j)}+ 2 E_{(i}\eta_{j)}\,\nonumber\\
    \delta \zeta^i=\,&-3\varepsilon^{ijk}\slashed{D}E_j\epsilon_k+\varepsilon^{ijk}\gamma\cdot \slashed{D}T_k\epsilon_j-4\gamma\cdot R(V)^i{}_{j}\epsilon^j-16i \gamma\cdot R(A)\epsilon^i-\frac{1}{2}D^i{}_{j}\,\nonumber\\
    \,&-\frac{3}{8}E^i\gamma\cdot T_j\epsilon^j+\frac{3}{8}E^j\gamma\cdot T_j\epsilon^j+ \frac{1}{8}E^jE_j\epsilon^i-4\Bar{\Lambda}_{L}\slashed{D}\Lambda_{R}\epsilon^i-4\Bar{\Lambda}_{R}\slashed{D}\Lambda_{L}\epsilon^i\,\nonumber\\ \,&-3\Bar{\Lambda}_R\slashed{D}\gamma_{ab}\Lambda_{L}\gamma^{ab}\epsilon^i+\frac{1}{2}\varepsilon^{ijk}\Bar{\Lambda}_{L}\gamma^a\epsilon_{j}\gamma_{a}\zeta_{k}-6\Bar{\Lambda}_{L}\Lambda_{L}\Bar{\Lambda}_{R}\Lambda_{R}\epsilon^i+\varepsilon^{ijk}\gamma \cdot T_j\eta_{k}\,\nonumber\\ \,& -3\varepsilon^{ijk}E_{j}\eta_{k}
\end{align}
The supercovariant curvatures associated with the gauge fields $e_{\mu}{}^{a}$, $\omega_{\mu}{}^{ab}$ and $\psi_{\mu}{}^i$, can be read off from their gauge transformations and are given as follows:
\begin{subequations}
\begin{align}
    R(P)_{\mu\nu}{}^{a}=\,&2\mathscr{D}_{[\mu}e_{\nu]}{}^a-\Bar{\psi}_{[\mu}{}^i\gamma^a \psi_{\nu]i}, \label{eq:RP3} \\
    R(M)_{\mu\nu}{}^{ab}=\,&2 \partial_{[\mu}\omega_{\nu]}{\!}^{ab}-2\omega_{[\mu}{\!}^{ab}\omega_{\nu] c}{\!}^b-4e_{[\mu}{\!}^{[a}f_{\nu]}{\!}^{b]} \nonumber \\
    &+\frac{1}{2}\Big(\Bar{\psi}^{i}_{[\mu}\gamma^{ab}\phi_{\nu]i}-2{\Bar{\psi}}_{[\mu i}\gamma_{\nu]}R(Q)^{ab i}-\frac{1}{4}\varepsilon_{ijk}\Bar{\psi}_{\mu i}\psi_{\nu j}T^{ab k}+\text{h.c}\Big), \label{eq:RM3} \\
    R(Q)_{\mu\nu}^{i}=\,& 2\mathscr{D}_{[\mu}\psi_{\nu]}^i-\gamma_{[\mu}\phi_{\mu]}{\!}^i-\frac{1}{8}\varepsilon^{ijk}\gamma \cdot T_{j}\gamma_{[\mu}\psi_{\nu]k}+\frac{1}{2}\varepsilon^{ijk}\Bar{\psi}_{[\mu j}\psi_{\nu]k}\Lambda_L\;, \label{eq:RQ3}
\end{align}
\end{subequations}
where, the derivative $\mathscr{D}_{\mu}$ is covariant with respect to all bosonic symmetries except the special conformal transformation. We need the variation of the curvature $R(Q)^i_{\mu\nu}$ for future sections, and it is given as below.
\begin{align}\label{RQ-trans-N3}
	\delta R(Q)_{ab}^{i}&=-\frac{1}{2}R(M)_{abcd}\gamma^{cd}\epsilon^{i}-\frac{1}{8}\varepsilon^{ijk}\left[\gamma\cdot T_{k}\gamma_{ab}+\frac{1}{3}\gamma_{ab}\gamma\cdot T_{k}\right]\eta_{j}\nonumber \\
	&\quad+\frac{1}{4}\Big[\gamma^{cd}\gamma_{ab}+\frac{1}{3}\gamma_{ab}\gamma^{cd}\Big]\Big[R(V)_{cd}{}^{i}{}_j\epsilon^{j}-\frac{i}{2}R(A)_{cd}\epsilon^{i}-\frac{1}{4}\varepsilon^{ijk}\slashed{D}T_{cd k}\epsilon_{j}\Big] 
\end{align}
The curvatures above satisfy the following constraints:
\begin{subequations}\label{eq:Constraints3}
\begin{align}
    R(P)_{\mu\nu}{}^{a} &= 0, \label{eq:RPZero3} \\
    e^{\nu}{}_{b}R(M)_{\mu\nu}{}^{ab} &= 0, \label{eq:RMZero3} \\
    \gamma^{\mu}R(Q)_{\mu\nu}^{i} &= 0. \label{eq:RQZero3}
\end{align}
\end{subequations}
One can solve the above constraints to obtain the dependent gauge fields $\omega_{\mu}^{ab}$, $f_{\mu}^a$ and $\phi_{\mu}^i$ in terms of the independent ones. We will need below the expression for the gauge field $f_\mu^a$ which is given below (up to fermionic terms which we have omitted).
\begin{align}\label{eq:FM2}
    f_{\mu}{\!}^a &= \frac{1}{2} R_{\mu}{\!}^{a}-\frac{1}{12}Re_{\mu}{\!}^{a}\;. 
\end{align}
In the above equations, we have used the following definitions:
\begin{align}
   R_{\mu}{}^{a}=e^{\nu}{}_{b}R_{\mu\nu}{}^{ab}\;, \;\;  R=e^{\mu}{}_{a}R_{\mu}{}^{a}
\end{align}
where $R_{\mu\nu}{}^{ab}\equiv \left(R(M)_{\mu\nu}{}^{ab}\right)_{|f=0}$. In the Poincar{\'e} gauge $b_\mu=0$, $R_{\mu\nu}{}^{ab}$, $R_{\mu}{}^{a}$ and $R$ have the interpretation of Riemann tensor, Ricci tensor and Ricci scalar respectively.

The $\mathcal{N}=3$ vector multiplet is an on-shell multiplet. It has been obtained in \cite{Hegde:2022wnb}  by rearranging the fields of the $\mathcal{N}=4$ vector multiplet. We summarize the field content in Table
\ref{tab:vector3}.
\begin{table}[H]
\small
    \centering
    \begin{tabular}{|c|c|c|c|c|
}
    \hline
         Field& Type & $SU(3)$ &$w$ &$c$ \\
         \hline
         $A_{\mu}$ & Boson & \textbf{1} &0&0\\ \hline
          $\xi_i$& Boson & ${\textbf{3}}$ & $1$& $-1$\\\hline
         $\psi_i$& Fermion & ${\textbf{3}}$ & $3/2$& $1/2$\\\hline
        $\theta_L$& Fermion & ${\textbf{1}}$ & $3/2$ & $3/2$\\
         \hline
    \end{tabular}
    \caption{$\mathcal{N}=3$ Vector multiplet}
    \label{tab:vector3}
\end{table}

The $Q$- and $S$-supersymmetry transformations for the fermions corresponding to $n$ abelian vector multiplets labeled by $I=1,2,\cdots,n$ are given by
\begin{align}
\delta \psi_i^{I}=\,&-\frac{1}{2}\gamma\cdot \mathcal{F}^{I +}\epsilon_{i}-2\varepsilon_{ijk}\slashed{D}\xi^{I k}\epsilon^j-\frac{1}{4}E_i\xi^{I k}\epsilon_{k}+\frac{1}{2}\Bar{\Lambda}_{L}\theta^{I}_{L}\epsilon_{i}\,\nonumber\\\,
&+\frac{1}{2}\gamma_a\epsilon^j\Bar{\Lambda}_R\gamma^a\Lambda_{L}\xi^{I k}\varepsilon_{ijk}+ 2\varepsilon_{ijk}\xi^{I j}\eta^{k}\,\nonumber\\\,
\delta \theta^{I}_{L}=\,&-2\slashed{D}\xi^{i}\epsilon_{i}-\gamma^a\Bar{\Lambda}_{L}\gamma_{a}\Lambda_{R}\xi^{I j}\epsilon_{j}+\frac{1}{4}\varepsilon_{ijk}E^i\xi^{I j}\epsilon^k-\Bar{\Lambda}_R\psi^{I}_i\epsilon^i-2\xi^{I i}\eta_{i}.
\end{align}
It was found in \cite{Hegde:2022wnb} that by coupling $3+n_V$ vector multiplets to the $\mathcal{N}=3$ Weyl multiplet one can obtain the matter coupled higher derivative $\mathcal{N}=3$ Poincar\'e supergravity. Writing the corresponding action in terms of Poincar\'e supergravity involves gauge fixing the additional conformal symmetries and solving the auxiliary field equations in a derivative expansion. However, for our work, we only need a few ingredients from this analysis which we outline below. In particular, we will provide a few of the gauge fixing conditions and auxiliary field equations of motion, without solving them in generality.

The additional symmetries that we need to gauge fix are the special conformal boosts, dilatations, $SU(3)$ and $U(1)$ R-symmetry transformations and $S$-supersymmetry transformations. The special conformal boosts are gauge fixed by setting,
\begin{align}\label{bmu-zero-N3}
b_\mu=0.
\end{align}
The dilatation gauge fixing condition is,
\begin{align}\label{D-gauge-N3}
\xi^{Ii}\xi_i^J\eta_{IJ}=-\frac{1}{\kappa^2}.
\end{align}
We also need the equation of motion for the auxiliary field $D^i{}_j$. The Lagrangian is schematically written as,
\begin{align}
\mathcal{L}=\mathcal{L}_{\text{V}}+\lambda\mathcal{L}_{\text{CSG}},
\end{align}
where $\lambda$ provides a control parameter for the derivative expansion while solving for the auxiliary fields. The $D^i{}_j$ equation of motion is then given as,
\begin{align}\label{D-eom-N3}
		\left[ \xi_i^I \xi^{Jj} \eta_{IJ}\right]_{\textbf{8}} &= \frac{\lambda}{3} D^j\;_i\ . 
\end{align}
We can therefore do the following decomposition,
    \begin{align}\label{D-decomp-N3}
		\xi_i^I \xi^{Jj} \eta_{IJ} &= - \frac{1}{3 \kappa^2}\delta_i^j + \frac{\lambda}{3} D^j\;_i, 
	\end{align}
where we have used \eqref{D-gauge-N3}. One needs to solve for $D^i{}_j$ in terms of the physical fields to construct the Lagrangian in Poincar\'e supergravity variables, and this induces a derivative expansion with $\lambda$ as the control parameter. However, for the leading order $O(\lambda^0)$, we have,
\begin{align}\label{D-leading-N3}
		\xi_i^I \xi^{Jj} \eta_{IJ} &= - \frac{1}{3 \kappa^2}\delta_i^j + \cO(\lambda^1), 
\end{align}
We can consider a $(3+n_V)\times (3+n_V)$ Hermitian matrix that encodes the scalars as,
\begin{align}
M^{IJ}\equiv \kappa^2\xi_i^I\xi^{Ji}.
\end{align}
The above matrix satisfies the relations,
\begin{align}
\text{Tr}(M\eta)&=-1,
\end{align}
due to \eqref{D-gauge-N3}. Further, at leading order, we have
\begin{align}\label{M-eta-M-N3}
M^\dagger \eta M=-\frac{1}{3}M,
\end{align}
due to \eqref{D-gauge-N3} and \eqref{D-leading-N3}. The relation \eqref{M-eta-M-N3} will be important for the analysis of fully supersymmetric solutions in the next section. We will show that the higher order terms are zero for the fully supersymmetric solution, thus allowing us to use the leading order relation for the matrix $M^{IJ}$.

\section{\texorpdfstring{$\mathcal{N}=3$ Fully Supersymmetric Solution}{\mathcal{N}=3 Fully Supersymmetric Solution}}\label{N3susy}
In this section, we find the fully supersymmetric solution in $\cN=3$ higher derivative Poincar\'e supergravity considered in the previous section. We will carry this out similar to the analysis of $\mathcal{N}=4$ supersymmetric solutions that we carried out in section-\ref{N4susy}. We will consider $S$-invariant combinations of fermions and set their variations to zero.

We need a compensating spinor for the $S$-supersymmetry transformation. The following combination of the fields transforms as a pure gauge field under $S$-supersymmetry:
\begin{align}
\Theta^{i}=\frac{1}{2}\xi^{-2}\Big(\varepsilon^{ijk}\xi_{j}^{I}\psi_{k}^{J}\eta_{IJ}-\xi^{I i}\theta_{R}^J\eta_{IJ}\Big), ~~\xi^2=\xi^{I j}\xi_{j}^{J}\eta_{IJ}
\end{align}
The $S$-supersymmetry transformation is 
\begin{equation}
    \delta_S \Theta^{i}=\eta^i
\end{equation}
This will allow us to construct $S$-invariant combinations of the fermions below. 

The Weyl multiplet fermion $\Lambda_L$ does not transform under $S$-supersymmetry. Therefore we can set its $Q$-supersymmetry transformation to zero directly.
\begin{align}
    \delta\Lambda_L=\,& -\frac{1}{4}E_{i}\epsilon^i+\frac{1}{4}\gamma.T_i\epsilon^i\overset{!}{=}0.
\end{align}
By collecting the terms according to their  $SU(3)$ and Lorentz representations, we obtain,
\begin{equation}
    E_i=0,~~T_{ab i}=0.
\end{equation}
With these conditions the $S$-supersymmetry transformation of the field $\chi_{ij}$ vanishes and its $Q$-supersymmetry transformation simplifies to,
\begin{equation}
    \delta \chi_{ij}=-8\varepsilon_{kl(i}\gamma.R(V)^l{}_{j)}\epsilon^k+\frac{1}{3}\varepsilon_{kl(i}D^l{}_{j)}\epsilon^k\overset{!}{=}0,
\end{equation}
which implies 
\begin{equation}\label{RV-D-zero-N3}
    R(V)_{\mu\nu}^i{}_{j}=0,~~D^i{}_{j}=0
\end{equation}
The second equation above is important as it simplifies the auxiliary field $D^i{}_j$ equation of motion \eqref{D-eom-N3} to \eqref{D-leading-N3}. Thus, for a fully supersymmetric solution \eqref{M-eta-M-N3} follows without needing any derivative expansion.

The vanishing variation of $\zeta^i$ implies 
\begin{equation}
    R(A)_{\mu\nu}=0.
\end{equation}
Finally, the variation of the gravitino field strength given in \eqref{RQ-trans-N3}, along with the above conditions, gives us the condition
\begin{equation}\label{Weyl-zero-N3}
    R(M)_{\mu\nu}{}^{ab}=0
\end{equation}
Now we turn to the variations of the vector multiplet fermions. We start with gaugino variations. We take the linear combination $(\psi_i^{I}-2\varepsilon_{ijk}\xi^{Ij} \Theta^k)$ and demand this to vanish under $Q$-supersymmetry.
\begin{align}
\,&(\delta\psi_i^{I}-2\varepsilon_{ijk}\xi^{Ij} \delta\Theta^k)\,\nonumber\\& = A^{I}{}_{ab}\gamma^{ab}\epsilon_{i}+B^{Ik}{}_{a}\gamma^a \varepsilon_{ijk}\epsilon^j+ C^{I j}{}_{iab}\gamma^{ab}\epsilon_j+D^{I jn}{}_{ia}\gamma^a \varepsilon_{jmn}\epsilon^n+E^{I jk}{}_{la}\gamma^a\varepsilon_{ijk}\epsilon^l\overset{!}{=}0
\end{align}
where 
\begin{align}
    A^{I}{}_{ab}=\,&-\frac{1}{2}(\mathcal{F}_{ab}^{+I}+\xi^{-2}\xi^{Ij}\xi^{K}_{j}\mathcal{F}_{ab}^{+L}\eta_{KL})\,\nonumber\\
    B^{Ik}{}_{a}=\,& -2(D_{a}\xi^{Ik}+\xi^{-2}\xi^{I l}\xi_{l}^{J}D_a\xi^{L k}\eta_{JL})\,\nonumber\\
    C^{I j}{}_{i ab}=\,& \frac{1}{2}\xi^{-2}\xi^{Ij}\xi_{i}^{K}\mathcal{F}_{ab}^{+L}\eta_{KL}\nonumber\\
    D^{I jn}{}_{ia}=\,& 2\xi^{-2}\xi^{I j}\xi^{K}_{i}D_a\xi^{L n}\eta_{KL}\nonumber\\
    E^{I jk}{}_{la}=\,&-2\xi^{-2}\xi^{Ij}\xi^{Pk}D_a\xi^Q_{l}\eta_{PQ}
\end{align}
We will consider in particular,
\begin{align}\label{AIab-N3}
A^{I}{}_{ab}=\,&-\frac{1}{2}(\mathcal{F}_{ab}^{+I}-M^{KI}\mathcal{F}_{ab}^{+L}\eta_{KL})=0,
\end{align}
where we have used \eqref{D-gauge-N3} in the units $\kappa=1$. We can contract the above with $M^{MN}\eta_{IM}$ and sum over the repeated indices to obtain,
\begin{align}\label{M-eta-F-N3}
\frac{2}{3}M^{KN}\eta_{KL}\mathcal{F}^{+L}_{ab}=0,
\end{align}
where we have used \eqref{M-eta-M-N3}. Note that we have argued below \eqref{RV-D-zero-N3} on how \eqref{M-eta-M-N3} holds for the fully supersymmetric solution at any order in the derivative expansion. Now, using \eqref{AIab-N3} and \eqref{M-eta-F-N3}, we obtain,
\begin{align}
\mathcal{F}_{ab}^{+I}=0.
\end{align}
We can similarly consider,
\begin{align}
 B^{Ik}{}_{a}=\,& -2(D_{a}\xi^{Ik}-M^{JI}D_a\xi^{L k}\eta_{JL})=0.
\end{align}
Upon using \eqref{M-eta-M-N3} this gives,
\begin{align}
D_a\xi^{Ii}=0.
\end{align}
From this, it follows that,
\begin{align}
M^{IJ}=\text{constants.}
\end{align}
Further analysis involving the derivative of the fermion $\psi_{i}^{I}$ gives
\begin{align}\label{kgauge-N3}
f_{\mu}^a=0.
\end{align}
The detailed arguments for the above two equations are completely analogous to how we obtained \eqref{mijcon} and \eqref{fmu0} in the case of $\mathcal{N}=4$ supergravity. Combining \eqref{kgauge-N3} above with \eqref{eq:FM2}, \eqref{Weyl-zero-N3} and \eqref{bmu-zero-N3}, we get,
\begin{align}
R_{\mu\nu\rho\sigma}=0.
\end{align}
Thus, we obtain the flat spacetime again as the unique fully supersymmetric solution in higher derivative matter coupled $\mathcal{N}=3$ Poincar\'e supergravity. The scalars can take arbitrary constant values, while the metric remains flat and the fluxes are zero. The result is summarized below in Table-\ref{tab:n3-flat}.

\begin{table}[H]
\small
    \centering
    \begin{tabular}{|c|c|}
    \hline
         Field& Full Supersymmetric solution\\
         \hline
         $F_{\mu\nu}^{I}$ & 0~(No flux)\\
         \hline
         $g_{\mu\nu}$&Flat spacetime\\
         \hline
         $M^{IJ}$&constant\\
         \hline
    \end{tabular}
    \caption{$\mathcal{N}=3$ Fully supersymmetric solution}
    \label{tab:n3-flat}
\end{table}

\section{\texorpdfstring{Reason for a richer supersymmetric vacua in $\cN=2$ supergravity}{Reason for a richer supersymmetric vacua in N=2 supergravity}}\label{N32}
$\cN=2$ Poincar\'e supergravity coupled to $n_V$ vector multiplets is gauge equivalent to $\cN=2$ conformal supergravity coupled to $(n_V+1)$ multiplets.
As discussed in the introduction, three fully supersymmetric solutions exist in  $\cN=2$ Poincar\'e supergravity. The non-stationary fully supersymmetric solutions are pp wave spacetimes. The fully supersymmetric stationary solutions include the  $AdS_2\times S^2$ geometry and flat spacetime. Even in the presence of higher derivative corrections, these two solutions remain the only fully supersymmetric stationary solutions in $\cN=2$ Poincar\'e supergravity. In contrast,  $\cN=4$ and $\cN=3$ supergravity have flat spacetime as the unique fully supersymmetric solution. 

We provide an argument for the difference in vacuum structures of these supergravity theories from the viewpoint of conformal supergravity. The most crucial equations in our analysis are the transformations of $\Lambda_i$ and $\Lambda_L$ in $\cN=4$ supergravity and $\cN=3$ supergravity, respectively. Since these fermions do not transform under $S$-supersymmetry, we do not need to add any $S$-supersymmetry compensator to analyze the corresponding Killing spinor equations. The $Q$-supersymmetry variation of these fermions sets the values of the auxiliary fields  $T_{ab}^{ij}$ and $T_{ab}^{i}$ to zero. These conditions force the electromagnetic flux to take zero values and moduli fields to take arbitrary constant values. As a consequence, the underlying geometry has to be necessarily flat.

One can construct $\cN$=2 matter-coupled Poincar\'e supergravity using the off-shell $\cN=2$ Weyl multiplet, the off-shell $\cN=2$ vector multiplet and on shell $\cN=2$ hypermultiplet. One can obtain these $\cN=2$ supermultiplets  via a supersymmetric truncation of the $\cN=3$ multiplets. Truncating the $\cN=3$
Weyl multiplet, one obtains the off-shell $\cN=2$ Weyl multiplet and an off-shell $\cN=2$ vector multiplet. Under a similar truncation, the on-shell $\cN=3$ vector multiplet  splits into an on-shell $\cN=2$ vector multiplet and an on-shell $\cN=2$ hypermultiplet. We do not provide the full truncation results here. For details, see \cite{Aikot:2024cne}. For our purpose, the following observation is important. In the truncation procedure, a set of fields is set to zero that would otherwise form an $\cN=2$ gravitino multiplet. This includes the S-invariant spinor $\Lambda_L$. As a consequence, there is no Killing spinor equation in $\cN=2$ supergravity, that would require us to set the auxiliary field $T_{ab}^{ij}$ to zero. In the Poincar\'e picture, the non-vanishing value of the $T_{ab}^{ij}$ field becomes the radius of the $AdS_2 \times S^2$ geometry and the moduli fields take constant values which are completely determined by the electromagnetic charges \cite{LopesCardoso:1998tkj}. 
\section{Conclusions and future directions}\label{conclusion}
In this paper, we used the framework of conformal supergravity to show that flat spacetime is the only fully supersymmetric solution in the matter coupled $\cN=4$ and $\cN=3$ higher derivative Poincar\'e supergravity that contains a class of higher derivative terms related to the Weyl squared term by supersymmetry. We have also argued that the vanishing of the auxiliary fields $T_{ab}^{ij}$ and $T_{ab}^i$ is the sole reason for getting flat spacetime as a 
unique fully supersymmetric solution in $\cN=4$ and $\cN=3$ supergravity respectively. Our results hold to all order in the derivative expansion for a class of supergravity Lagrangians that are constructed using the conformal supergravity Lagrangian of \cite{Butter:2019edc,Hegde:2022wnb} followed by gauge fixing the additional super-conformal symmetries. However, the input provided by the details of the conformal supergravity action for our results is very limited. In particular, only the $D$-field equation of motion has been used from the conformal supergravity action to arrive at our results. Therefore we expect our results to hold more generally for $\cN=3,4$ Poincar\'e supergravity to all order in the derivative expansion. The class of higher derivative theories considered in this paper are the most general ones within the framework of $\cN=4$ and $\cN=3$ conformal supergravity theories constructed using the standard Weyl multiplet. However, one may be able to construct more general class of higher derivative theories within the framework of $\cN=4$ and $\cN=3$ conformal supergravity but instead using the dilaton Weyl multiplets \cite{Ciceri:2024xxf,Adhikari:2024esl,Adhikari:2024qxg,Adhikari:2025wwb}. It would be interesting to extend our analysis to such class of theories.

Our analysis implies that the horizon of the extremal black holes
in $\cN=4$ supergravity can not be fully supersymmetric. This aligns with a recent work \cite{Chen:2024gmc}, where the authors have argued using super Schwarzian theory and superconformal groups, that there is no decoupled $AdS_2$ near horizon geometry that preserves all the $16$ supercharges describing a heterotic string. This situation is in contrast to $\cN=2$ supergravity, where the near horizon geometry of extremal black holes is fully supersymmetric $AdS_2 \times S^2$ geometry and near horizon field configuration is completely fixed by the black hole charges. This is encoded in the so-called attractor mechanism. The full extremal black hole solutions of $\cN=2$ supergravity possess residual $\cN=1$ supersymmetry and full $\cN=2$ supersymmetry are restored as one approaches the horizon as well as asymptotic infinity. These solutions come under the IWP class of solutions, and when embedded in $\cN=4$ supergravity, they always preserve $1/4$-of the supersymmetries \cite{Bergshoeff:1996gg}. The general stationary solutions in two-derivative $\cN=4$ supergravity are known as SWIP solutions, which are described by two complex harmonic functions, and they contain all the $\cN=2$ IWP class of solutions. The near-horizon geometry of $1/4$-supersymmetry preserving SWIP solutions, are typically $AdS_2 \times S^2$ geometry. The higher derivative correction to these $\cN=4$ SWIP solutions are not known and it would be interesting to find such corrections and analyze the amount of supersymmetries preserved by such solutions. Further, since $AdS_2 \times S^2$ geometry is fully supersymmetric in an $\cN=2$ theory, one may expect that it would preserve $1/2$ of the 16 supersymmetries in higher derivative $\cN=4$ supergravity. A study of $1/2$-BPS solutions of higher derivative $\mathcal{N}=4$ supergravity from an off-shell approach analogous to the one taken in \cite{LopesCardoso:2000qm} is therefore of interest. It will be interesting to see how this analysis will align with recent work on the number of supercharges that can be preserved by a BPS black hole using a super-Schwarzian approach \cite{Heydeman:2025vcc}. We hope to report on these issues in the future.

Classifying supersymmetric solutions of four-dimensional $\cN=2$ and $\cN=4$ supergravity has been done with great effort over the last 40 years. Whereas an understanding of the higher derivative analysis has been achieved for the former, the same is lacking for the latter. Consequently, higher derivative corrections to the entropy of black holes in $\cN=4$ supergravity have been calculated using the $\cN=2$ black hole entropy formula \cite{LopesCardoso:2006ugz, Mohaupt:2008gt}. For example, `small black holes' have zero horizon area in the two-derivative $\cN=4$ supergravity and it is expected to develop the horizon area once higher derivative corrections are taken into account \cite{Sen:1995in,LopesCardoso:2006ugz, Dabholkar:2004dq, Dabholkar:2005dt, Mohaupt:2008gt}. There is an alternative claim that thermally, there is a transition to a gas of free strings before reaching any BPS small black hole configuration \cite{Chen:2021dsw, Chen:2024gmc}.  It will be interesting to study these solutions in higher derivative $\cN=4$ supergravity directly rather than using truncated $\cN=2$ supergravity formalism . In this context, it is essential to classify the supersymmetric solutions of the higher derivative $\cN=4$ supergravity. The present work is an initial step towards that goal. 

\acknowledgments

We thank Ashoke Sen and Amitabh Virmani for discussions. AB and BS thank Harish
Chandra Research Institute, Prayagraj for its hospitality during the initial stages of this
project. AB thanks organizers of National Strings Meeting 2024, IIT Ropar where this work
was presented. SH is funded by the European Union (ERC, UNIVERSE PLUS, 101118787).
BS thanks the organisers of the workshop “New Frontiers in Supergravity, Strings and Related
Topics” held at Hefei, China, where this work was presented. Views and opinions expressed
are however those of the authors only and do not necessarily reflect those of the European
Union or the European Research Council Executive Agency. Neither the European Union
nor the granting authority can be held responsible for them.

\bibliography{bibliography}
\bibliographystyle{JHEP}

\end{document}